\documentclass[aps,prl,showpacs,twocolumn]{revtex4}
\usepackage{amssymb}
\usepackage{graphicx}

\begin{document}

\title{The difference between Si \ and Ge(001) surfaces in the initial stages of
growth}
\author{G. Brocks, J.H. Snoeijer, P. J. Kelly, H. J. W. Zandvliet, and Bene Poelsema}
\affiliation{Faculty of Applied Physics and MESA+ Research Institute, University of
Twente, P. O. Box 217, 7500 AE Enschede, The Netherlands.}

\date{\today}

\begin{abstract}
The initial stages of growth of Ge and Si on the Ge(001) surface are studied
and compared to growth on the Si(001) surface. Metastable rows of diluted
ad-dimers exist on both surfaces as intermediate stages of epitaxial growth.
Unexpectedly, for Ge(001) these rows are found exclusively in the $%
\left\langle 310\right\rangle $ directions, whereas on Si(001) the preferred
direction is $\left\langle 110\right\rangle $. This qualitative difference
between Si and Ge surfaces reflects the subtle difference in the chemistry
of these two elements, which has direct consequences for epitaxial growth on
these surfaces.
\end{abstract}


\maketitle

The Si(001) surface is probably the most widely studied semiconductor
substrate for understanding the processes which govern epitaxial growth,
both experimentally and theoretically. Scanning tunneling microscopy (STM)
has revealed a surprising richness of adsorbed structures even in the
initial stages of homoepitaxial growth. In recent years growth of Si/Ge
alloys and multilayers has received increased attention as such materials
find new applications in semiconductor devices. For example, the controlled
growth of these materials allows for band gap engineering in which,
depending upon the composition, the band gap can be varied between that of
Ge and Si. Growth on the Ge(001) surface has received much less attention
than that on Si(001) which is surprising since in heterostructures both
these surfaces play an equally important role. Probably this lack of
attention reflects the general perception that one can derive the properties
of the Ge surface from those of the Si surface because the chemistry of Si
and Ge is very similar. Indeed Si and Ge(001) surfaces do have the same
basic reconstruction and some of the structures formed by adsorbed species
are very similar. As we will show in this paper, however, growth on Ge(001)
does \emph{not} necessarily follow the same pathway as on Si(001), despite
the similarity in chemical bonding. In particular, the so-called
\textquotedblleft dilute\textquotedblright\ rows of ad-dimers, which form
crucial intermediate structures in the growth process, have a different
structure and dynamical behavior. This is caused by the difference in
reactivity of the Ge and Si surfaces and it results in a much more ordered
growth on the Ge surface.

\begin{figure}
\includegraphics[width=6.5cm,keepaspectratio=true,clip=true]{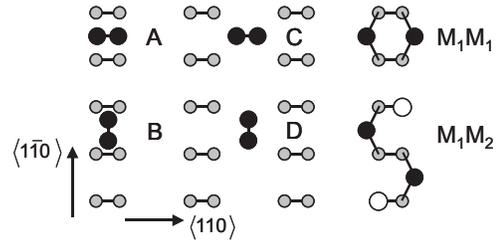}
\caption{ Structures of Si or Ge ad-dimers and adatom
pairs on Si or Ge(001) surfaces. Gray dumbbells represent substrate dimers,
black circles the adsorbed atoms, and white circles the surface dangling
bonds associated with the $M_1M_2$ structure.}
\label{struct}
\end{figure}

Similar to Si(001), the basic reconstruction of the Ge(001) surface consists
of surface atoms forming dimers which are arranged in rows. At room
temperature, adsorbed single Si or Ge atoms are very mobile on the surface
and the smallest structures observed are ad-dimers, i.e. bonded pairs of
adatoms. Such ad-dimers play a prominent role in the low temperature growth
of Ge and Si, and much work has been devoted recently to studying their
energetics and their mobility \cite{Zandvliet97,Zandvliet00}. The basic
(metastable) structures of ad-dimers are shown schematically in Fig. \ref
{struct}. Most of these dimer structures prove to be accessible kinetically
and they can be observed in STM. From a population analysis one can, in
principle, establish their relative energies. However, because of the
kinetic barriers involved in the various diffusion processes, it can in
practice be very difficult to achieve equilibrium on a surface. 

It is also feasible to obtain such energy differences from first principles
density functional calculations, simultaneously optimizing both the
electronic structure and the geometry \cite{Brocks96,Calcnote}. The
calculated total energy differences for Ge/Si ad-dimers on Ge/Si(001)
surfaces are given in Table \ref{en_diff}. In detail the results depend somewhat upon the
computational parameters used \cite{Brocks96,Calcnote,Yamasaki96,Smith96,Khare99},
but the overall trend is unambiguous. If one compares the numbers in Table 
\ref{en_diff}, one observes that the energy ordering is roughly the same in
all cases. The $B$-type structure on-top of a substrate dimer row is lowest
in energy. The ``on-top'' $A$ and ``in-between'' $C$ structures are somewhat
higher in energy, the on-top position being slightly more favorable. The
``in-between'' $D$ structure is substantially higher in energy, as are the
adatom pairs $M_{1}M_{1}$ and $M_{1}M_{2}$. One would conclude that chemical
bonding is the same for corresponding structures on the Ge(001) and Si(001)
surfaces.

\begin{table}
\begin{center}
\begin{tabular}{ccccc}
\toprule
& Ge/Ge(001) & Si/Ge(001) & Ge/Si(001) & Si/Si(001) \\ 
\colrule
$A$ & 0.18 & 0.24 & 0.06 & 0.01 \\ 
$C$ & 0.23 & 0.26 & 0.19 & 0.28 \\ 
$D$ & 0.80 & 0.63 & 1.01 & 0.95 \\ 
$M_{1}M_{1}$ & 0.48 & 0.76 & 0.27 & 0.39 \\ 
$M_{1}M_{2}$ & 0.65 & 0.93 & 0.53 & 0.71 \\
\botrule
\end{tabular}
\end{center}
\caption{Energies (eV) of the structures shown in Fig.~\protect\ref{struct} relative
to the lowest energy structure, which is $B$ in all cases.}
\label{en_diff}
\end{table}

From the similarity of the energy landscapes sampled by diffusing species
on these surfaces one would draw a similar conclusion. Diffusion barriers
are quite hard to obtain from calculations, since one has to guess complete
diffusion paths. Diffusion barriers are more easily obtained from STM
experiments by timing the intervals between hopping events and extracting
jump rates. For Ge ad-dimers on Ge(001), jump rates at room temperature have
been measured for ad-dimers \cite{Galea00}. These can be
translated into a diffusion barrier of $0.83$ eV for diffusion of 
ad-dimers parallel to the substrate dimer rows (i.e. in the $\langle %
1\overline{1}0\rangle $ direction, cf. Fig. \ref{struct}) and $0.95$ eV for
diffusion across the rows (in the $\langle 110\rangle $ direction). The
corresponding published numbers for diffusion of Si ad-dimers on the Si(001)
surface are (in eV): $0.94\pm 0.09$ \cite{Swarzentruber96}
or $1.09\pm 0.05$ \cite{Borovsky97} (parallel); $1.36\pm 0.06$ \cite{Borovsky97} (across). 
The diffusion barriers on Ge(001) are
lower, and the anisotropy of the diffusion is smaller. The former can be
related to the lower cohesive energy of Ge as compared to Si, the latter can
be explained by a subtle difference in surface structures. Whereas the bulk
lattice constant of Ge is 4\% larger than that of Si, the bond length of the
surface dimers is 9\% larger. It is then reasonable that the energy
landscape for a diffusing ad-dimer on the Ge(001) surface is flatter, since
the ad-dimer can more easily bridge the troughs between the substrate dimer
rows. That the smaller anisotropy for diffusion can be attributed to the
Ge(001) substrate rather than to the ad-dimer is confirmed by the diffusion
barriers of a Si ad-dimer on the Ge(001) surface, which are $0.83\pm 0.05$,
and $1.0$ eV for diffusion along and across the substrate dimer rows \cite
{Zoethout98}. Skipping the quantitative differences, one notices that the
relative order of the diffusion barriers in different directions is the same
on both surfaces. This indicates a similar energy landscape for the
diffusing species, at least with respect to the critical points of that
landscape.

\begin{figure}
\includegraphics[width=7.5cm,keepaspectratio=true,clip=true]{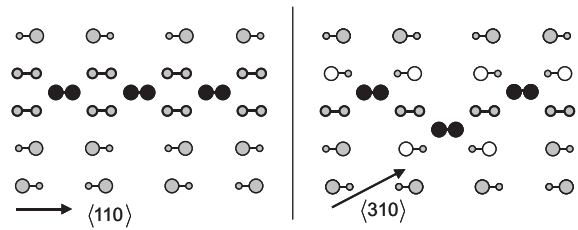}
\caption{ Schematic representation of adsorbed $\langle
110 \rangle$ and $\langle 310 \rangle$ rows, as in Fig. \ref{struct}. The
size of the circles represents the height of the corresponding atoms.}
\label{calcrows}
\end{figure}

The theoretical and experimental evidence presented so far supports the
suggestion that Ge(001) behaves similarly to Si(001); ad-dimers have similar
structures and energies and their diffusion processes are similar. Since
ad-dimers form the nuclei for growth on the surface, it is not unreasonable
to suggest that growth on the Ge(001) and Si(001) surfaces proceeds via the
same pathway. However, a distinct \emph{difference} emerges in the next
step. From STM studies of growth on the Si(001) surface, chain-like
structures of ad-dimers in the $\left\langle 110\right\rangle $ direction
have been identified that are perpendicular to the substrate dimer rows \cite
{Bedrossian95,Vanwingerden97,Qin97}. These are shown schematically in Fig. 
\ref{calcrows}. They are observed both for Si on Si(001) as well as for Ge
on Si(001) and are termed \textquotedblleft dilute dimer
rows\textquotedblright . These are generally believed to play an important
role as intermediate structures in the growth process \cite{Qin97}.

\begin{figure}
\includegraphics[width=6.5cm,height=9.57cm,clip=true]{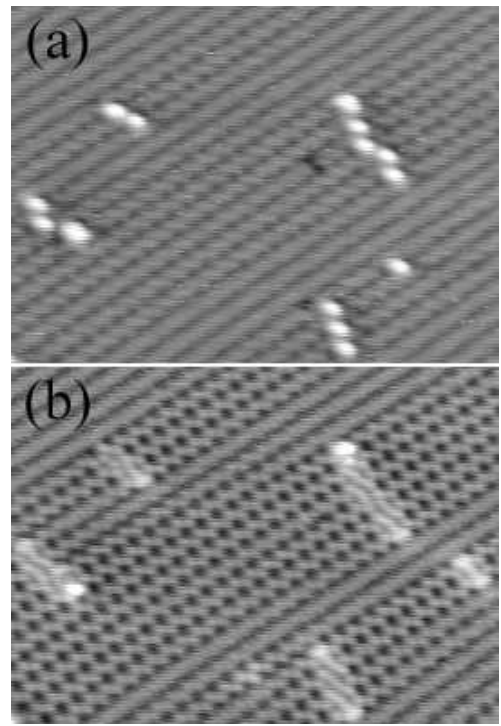}
\caption{ STM images (size 13.5 nm x 17 nm) of a
Ge(001)surface after the deposition of 0.02 ml Ge. (a) Empty state image;
sample bias 1.6 V, tunneling current 0.7 nA. (b) Filled state image; sample
bias $-1.6$ V, tunneling current 0.7 nA. }
\label{dimerrows}
\end{figure}

On Ge(001) such dilute dimer rows are \emph{not} observed at all, neither
after deposition of Si nor of Ge. Instead the structures which are most
frequently observed after low temperature deposition are rows of ad-dimers
which are oriented in the $\left[ 310\right] $ direction. Because of
symmetry, ad-dimer rows in the $\left[ 130\right] $ direction have the same
energy, and \textquotedblleft zigzag\textquotedblright\ rows which consist
of $\left[ 310\right] $ segments alternating with $\left[ 130\right] $
segments occur very often, cf. Figs. \ref{calcrows} and \ref{dimerrows}. In
the following we will use the phrase \textquotedblleft $\left\langle
310\right\rangle $ row\textquotedblright\ to label all such structures. $%
\left\langle 310\right\rangle $ rows are also observed on Si(001), but
gentle annealing of the surface seems to reduce their number in favor of the
number of the dilute rows in the $\left\langle 110\right\rangle $ direction.
This would suggest that on Si(001) the $\left\langle 110\right\rangle $ row
is thermodynamically more stable, whereas the $\left\langle 310\right\rangle 
$ row is easily accessible kinetically \cite{Vanwingerden97,Qin97}. On
Ge(001) gentle annealing to 400 K renders the segments of a $\left\langle
310\right\rangle $ row mobile, but a conversion into a $\left\langle
110\right\rangle $ row is never observed \cite{Galea00}.

As discussed above it is very difficult to achieve thermal equilibrium on a
semiconductor surface. Moreover, annealing can only be done in a very gentle
way, since increasing the temperature too much would destroy the absorbed
\textquotedblleft dilute\textquotedblright\ structures in favor of epitaxial
islands. The stability of dilute dimer rows can be checked by first
principles calculations. Table \ref{dilendiff} gives the relative energies
of dilute dimer rows of Si and Ge on Si(001) and Ge(001) substrates. Listed
are the energy differences $E_{\left\langle 310\right\rangle
}-E_{\left\langle 110\right\rangle }$ per adsorbed dimer between the $%
\left\langle 310\right\rangle $ and $\left\langle 110\right\rangle $
ad-dimer rows in the \ geometries shown in Fig. \ref{calcrows}. The results
are obtained for infinite rows which are compatible with the $p(4\times 4)$
symmetry of the surface supercell used in the calculations \cite{Calcnote}.
The energy of a straight $\left\langle 310\right\rangle $ row would probably
be somewhat lower than that of the zigzag row consisting of alternating $%
\left[ 310\right] $ and $\left[ 130\right] $ segments shown in Fig. \ref
{calcrows}, and one would expect the energy of the latter to be an upper
bound for \textquotedblleft $\left\langle 310\right\rangle $%
\textquotedblright -like rows. Even so, one can draw an unambiguous
conclusion. On the Si(001) surface, $\left\langle 110\right\rangle $ dilute
ad-dimer rows are lowest in energy by 0.2-0.3 eV/ad-dimer, whereas on
Ge(001) the oblique or zigzag $\left\langle 310\right\rangle $ rows are
lowest in energy by 0.1-0.3 eV/ad-dimer. This conclusion holds both for Si
and Ge ad-dimer rows, so the most stable structure seems to be dictated by
the \emph{surface} rather than by the \emph{adsorbed species}.

\begin{table}
\begin{center}
\begin{tabular}{ccccc}
\toprule
& Ge/Ge(001) & Si/Ge(001) & Ge/Si(001) & Si/Si(001) \\ 
$\Delta E$(eV) & $-0.12$ & $-0.28$ & $+0.28$ & $+0.18$ \\
\botrule
\end{tabular}
\end{center}
\caption{Energy difference $\Delta E=E \langle 310 \rangle - E \langle 110
\rangle $ per ad-dimer between rows of the geometries shown in Fig.~\ref{calcrows}}
\label{dilendiff}
\end{table}

In the following we will rationalize these results. The Si(001) and Ge(001)
surface structures with lowest energy both have $c(4\times 2)$ periodicity
corresponding to the arrangement of buckled surface dimers shown
schematically in Fig. \ref{calcrows}. Adsorption of an ad-dimer row
introduces a \textquotedblleft domain wall\textquotedblright\ between two $%
c(4\times 2)$ domains, which can be observed in Fig. \ref{calcrows} by
noting that the substrate dimer rows beneath the adsorbed row are forced to
be symmetric (non-buckled) \cite{Qin98}. This causes strain in the
substrate. A $\left\langle 310\right\rangle $ ad-dimer row introduces only
one symmetric row in the substrate, whereas a $\left\langle 110\right\rangle 
$ ad-dimer row introduces two such symmetric substrate rows. From the
induced distortion of the surface geometry one would conclude that $%
\left\langle 310\right\rangle $ ad-dimer rows are more favorable than $%
\left\langle 110\right\rangle $ ad-dimer rows, since the former introduce
fewer symmetric dimers and thus less strain in the substrate. On a clean
surface, symmetric dimers are also found to be higher in energy than buckled
dimers \cite{Ramstad95}. Moreover, the energy difference between symmetric
and buckled surface dimers is larger on the Ge(001) surface than on the
Si(001) surface \cite{Krueger95}. Assuming this also holds for surface
dimers underneath an adsorbed row, one would expect $\left\langle
310\right\rangle $ ad-dimer rows on the Ge(001) surface to be relatively
more stable.

We have not yet considered the chemical bonding between substrate and
adsorbed dimers. A row of ad-dimers in the $\left\langle 310\right\rangle $
structure leaves dangling bonds on the substrate. These are indicated in
Fig.~\ref{calcrows}; simple electron counting would assign one electron per
dangling bond. Being (partially) filled states with a high energy, such
dangling bonds show up very prominently in a filled state STM image, cf.
Fig. \ref{dimerrows}. In fact, they completely mask the adsorbed dimers in
the filled state image. In contrast, a row of ad-dimers in the $\left\langle
110\right\rangle $ structure leaves no such dangling bonds. On the basis of
\ simple chemical reasoning one would expect the $\left\langle
110\right\rangle $ ad-dimer rows to be more stable than the $\left\langle
310\right\rangle $ ad-dimer rows, since the former leads to a smaller number
of dangling bonds. Considering the two factors mentioned in this and the
previous paragraph, the stability of the $\left\langle 110\right\rangle $
versus $\left\langle 310\right\rangle $ ad-dimer row structures results from
a competition between the penalty for forming dangling bonds on the
substrate, which favors the former structure, and the penalty for forming
symmetric substrate dimers, which favors the latter structure.

The Si(001) surface shows a strong tendency to minimize the number of
dangling bonds on the substrate \cite{Brocks93}. On a clean surface the
surface dimers are bonded by $\pi $-bonds, as well as by $\sigma $-bonds.
Adsorbed Si/Ge adatoms or single and ``dilute'' ad-dimers break these $\pi $%
-bonds but leave the $\sigma $-bonds intact. If only one atom of a substrate
dimer is involved in the bonding to an adsorbed atom, the other atom of that
dimer is left with a dangling bond, which is the remnant of the broken $\pi $%
-bond. One expects such a dangling bond to be higher in energy than a $\pi $%
-state and thus be more reactive. This is illustrated by the energy
difference between the $M_{1}M_{2}$ and the $M_{1}M_{1}$ structures, Table 
\ref{en_diff}, which can roughly be interpreted as the penalty for creating
two dangling bonds on the substrate, Fig. \ref{struct}. This creation energy
is 0.26-0.31 eV per pair of dangling bonds on a Si(001) surface, as compared
to only 0.17 eV on a Ge(001) surface. Apparently $\pi $-bonding is stronger
on Si(001) than on Ge(001) \cite{Pibond}.

An adsorbed $\left\langle 110\right\rangle $ row of ad-dimers only leaves
dangling bonds on the surface at the end of a row, whereas a $\left\langle
310\right\rangle $ row gives rise to two dangling surface bonds per adsorbed
dimer. Chemical bonding thus favors adsorption in the $\left\langle
110\right\rangle $ direction. On the basis of chemical bonding alone one
would estimate the energy difference between the $\left\langle
310\right\rangle $ and $\left\langle 110\right\rangle $ rows on Si(001) to
be $E(M_1M_2) - E(M_1M_1) \sim 0.3$ eV/ad-dimer in favor of the latter. The
calculated energy differences of \ Table \ref{dilendiff} for adsorbed rows
on Si(001) are smaller, but not by much. This indicates that the strain
induced by the $\left\langle 110\right\rangle $ row is (at most) $0.1$
eV/ad-dimer higher than that induced by the $\left\langle 310\right\rangle $
row. A similar strain difference on the Ge(001) surface would already to a
large part counteract the small energy difference between $\left\langle
310\right\rangle $ and $\left\langle 110\right\rangle $ adsorbed rows of $%
<0.2$ eV/ad-dimer calculated on the basis of chemical bonding. In fact, from
the calculated energy differences of \ Table \ref{dilendiff} for adsorbed
rows on Ge(001) one would conclude that the strain induced by the $%
\left\langle 110\right\rangle $ row is $\gtrsim 0.3$ eV/ad-dimer higher than
that induced by the $\left\langle 310\right\rangle $ row. Note that this
number compares very well with energy difference between symmetric and
buckled dimers on the clean Ge(001) surface \cite{Krueger95}. Energetically
the strain difference on the Ge(001) surface tips the balance in favor of \ $%
\left\langle 310\right\rangle $ adsorbed rows as compared to $\left\langle
110\right\rangle $ rows.

We speculate that the difference in intermediate structures, $\left\langle
110\right\rangle $ vs. $\left\langle 310\right\rangle $, found on the
Si(001) and Ge(001) surfaces has a substantial influence on the epitaxial
growth on these surfaces. The effective binding energy between ad-dimers in
a $\left\langle 110\right\rangle $ row on the surface can be estimated by $%
E_{\mathrm{bond}}=\frac{1}{2}(E_{\left\langle 110\right\rangle }-2E_{\mathrm{%
\dim }}+E_{\mathrm{surf}}),$ from the total energies $E_{\left\langle
110\right\rangle }$ of the adsorbed row, $E_{\mathrm{\dim }}$ of the
adsorbed isolated dimer and $E_{\mathrm{surf}}$ of the clean surface,
respectively. This estimate gives quite a high binding energy of $E_{\mathrm{%
bond}}\sim 0.4$ eV for Si and Ge $\left\langle 110\right\rangle $ ad-dimer
rows on Si(001) while other rows have a binding energy of $E_{\mathrm{bond}%
}\lesssim 0.2$ eV.

Several mechanisms have been proposed by which a $\left\langle
110\right\rangle $ (dilute) ad-dimer row is transformed into an epitaxial
structure. The density of dimers in an epitaxial row is twice as high as in
a dilute row, so the transformation has to involve the incorporation of
additional adatoms or dimers and/or the collapse of part of a dilute row
into an epitaxial one \cite{Brocks96,Vanwingerden97}. In any case the dilute 
$\left\langle 110\right\rangle $ row acts as a growth nucleus and the
epitaxial structure is also formed in the $\left\langle 110\right\rangle $
direction. The transformation by insertion of adatoms/dimers can take place
simultaneously at several positions in the dilute row. If these insertions
take place in an uncorrelated way, this can easily lead to so-called
\textquotedblleft missing dimer\textquotedblright\ defects \cite{Bowler95},
i.e. epitaxial rows in which one or more dimers are missing. Such defects
are quite common on the Si(001) surface and are extremely difficult to get
rid off, even after careful annealing.

Ge ad-dimers in a $\left\langle 310\right\rangle $ row on Ge(001) are much
less strongly bound (the calculated effective binding energy is $E_{\mathrm{%
bond}}$ $\sim 0.1$ eV) and one expects such rows to break up easily.
Moreover, even in a $\left\langle 310\right\rangle $ row the Ge ad-dimers
are quite mobile \cite{Galea00}. Therefore we propose that the $\left\langle
310\right\rangle $ row imposes less restriction on epitaxial growth on
Ge(001) than does the $\left\langle 110\right\rangle $ row on Si(001) and
that it does not promote the formation of missing dimer defects. The result
is that epitaxial growth on the Ge(001) surface can occur almost defect-free
and that the Ge(001) surface contains far fewer defects than the Si(001)
surface.


\begin{thebibliography}{}
\bibitem{Calcnote}  Our supercell comprises a 10 layer slab of Ge (or
Si), a vacuum region of 16.3 \AA\ and a $p(4\times 4)$ surface cell. The
atomic positions are optimized, constrained by inversion symmetry through
slab center, and fixing the middle two layers in
their bulk positions. Pseudopotentials are taken from Ref. \cite{FH90}. A
plane wave kinetic energy cutoff of 9 Ry and a surface Brillouin zone
sampling of 2 $\mathbf{k}$-points are used. Increasing the number of $%
\mathbf{k}$-points to 4 changes the energies in Table \ref{en_diff} by $\sim
0.02$ eV or less; increasing the cutoff has a similar effect. A standard
local density functional is used, cf. Ref. \cite{Brocks96}; using the
gradient corrected PW91 functional changes the energies by $\sim
0.02 $ eV, except for that of the $D$ structure, which increases by $\sim
0.15$ eV.
\bibitem{Pibond}  On an absolute energy scale $\pi $-bonding in both Si and
Ge is of course weak compared to that in first row elements like C, cf. Ref. 
\cite{Krueger95}.
\end{thebibliography}
\end{document}